# Pilot study of electrochemical reduction of selected nucleotides and double-stranded DNA at pristine micro-/ultrananocrystalline boron-doped diamond electrodes at very negative potentials

Michal Augustín[1*], Vlastimil Vyskočil[2], Ondrej Szabó[1], Kateřina Aubrechtová Dragounová[1], Rene Pfeifer[1], Frank-Michael Matysik[3], Jiří Barek[2], Marián Marton[4], Alexander Kromka[1]

[1]Institute of Physics, Czech Academy of Sciences, Cukrovarnická 10/112, 162 00, Prague 6, Czech Republic

[2]Charles University, Faculty of Science, Department of Analytical Chemistry, UNESCO Laboratory of Environmental Electrochemistry, Albertov 6, 128 43 Prague 2, Czech Republic

[3]University of Regensburg, Faculty of Chemistry and Pharmacy, Institute of Analytical Chemistry, Chemo- and Biosensors, Universitätsstraße 31, 930 40 Regensburg, Germany

[4]Institute of Electronics and Photonics, Faculty of Electrical Engineering and Information Technology, Slovak University of Technology in Bratislava, Ilkovičova 3, 841 04 Bratislava, Slovakia

* Corresponding author. *E-mail address:* augustin@fzu.cz

Abstract

Pristine polycrystalline boron-doped diamond electrodes (BDDEs) – microcrystalline (B-MCDE) and ultrananocrystalline (B-UNCDE) were applied for the electrochemical reduction of several selected purine nucleotides – Guanosine 5'-monophosphate (GMP), 2'-Deoxyguanosine 5'-monophosphate (dGMP), Adenosine 5'-monophosphate (AMP), Adenosine 5'-diphosphate (ADP), Adenosine 5'-triphosphate (ATP), and pyrimidine nucleotides – Cytidine 5'-monophosphate (CMP), Thymidine 5'-monophosphate (TMP), as well as low-molecular-weight double-stranded DNA (dsDNA) at very negative potentials *via* linear sweep voltammetry (LSV). Three different types of electrode surfaces were employed – "H-terminated" B-MCDE (H-B-MCDE) and "O-terminated" B-MCDE/B-UNCDE (O-B-MCDE/O-B-UNCDE). It was found that electrochemical reduction of all tested analytes (except GMP) is possible at H-B-MCDE. On the other hand, electrochemical reduction of all selected analytes (except dsDNA) is possible at O-B-MCDE/O-B-UNCDE. Ambient oxygen and preadsorption step in the manner of incubation of the corresponding sensor in the analyte solution for 30 seconds had a profound effect on the repeatability of the results and, in the case of H-B-MCDE, also on the magnitude of voltammetric signals and the possibility of electrochemical reduction of dGMP. The use of the proposed sensors and their main advantages and disadvantages for the voltammetric determination of the tested analytes (demonstrated *via* electrochemical reduction of AMP) in the presence of hydrogen evolution reaction (HER) is also discussed.

1. Introduction

Since their introduction in 1992, boron-doped diamond electrodes (BDDEs) have been established as a fascinating material for various applications in different fields of electrochemistry [1-11]. Concerning the electroanalysis, BDDEs are characterised by various unique properties (wide potential window up to 3.5 V in aqueous solutions depending on boron content, surface pretreatment – O- or H-termination, high signal-to-noise ratio (S/N), low background currents, resistance to surface passivation, mechanical and chemical stability, and biocompatibility [5, 12-14, 25]. Moreover, BDDEs can be used for the destruction of organic substances in wastewater treatment *via* anodic generation of hydroxyl radicals [5, 15-24]. BDDEs also retain some of the remarkable properties of the diamond, such as hardness, thermal conductivity, and remarkable chemical inertness [5, 25]. Their application in electroanalysis, surface pretreatment, and electrochemical activation of BDDEs is highly important and needs to be optimised for electrochemical sensing. BDDE pretreatment and activation procedures frequently rely on two electrochemical procedures – anodic pretreatment resulting in O-terminated BDDE (O-BDDE) or cathodic pretreatment resulting in H-terminated BDDE (H-BDDE) [26].

Anodic pretreatment incorporates oxygen atoms at the BDDE surface, e.g., *via* carbon reaction with •OH radicals generated in aqueous media of pH<9.0. Oxygen-containing groups are thus formed (e.g., hydroxy, carbonyl, carboxyl, and ether groups) by applying highly positive currents densities or potentials (above +2.0 V; in the region of water decomposition reaction) from a few seconds to minutes (often directly in the analyte solution as a surface renewal/activation step between individual measurements). Cathodic pretreatment can be achieved by exposure of the BDDE surface to very negative potentials (more negative than -2.0 V in the region of hydrogen evolution reaction – HER) or negative current densities [26-33]. Prediction of the preferable pretreatment procedure for particular analytes is generally difficult, and both anodic and cathodic procedures are frequently examined. From the practical point of view, the cathodic pretreatment is considered less favourable, as it has to be applied right before the proper experiments to ensure reliable and reproducible results, mainly when the corresponding BDDE has not been used for a longer period of time due to instability of H-terminated surfaces exposed to air oxygen [26, 30-33].

The electrochemical activity of nucleic acids (NAs) at various electrode surfaces is predominantly attributed to the electrochemical activity of its components, nucleobases, and sugar residues. In general, various DNA components are electrochemically reducible at mercury-based electrodes (HMDE or non-toxic silver amalgam electrodes), whereas all types of nucleobases can be electrochemically oxidised at various carbon-based electrodes [34-35]. Moreover, the introduction of pyrolytic graphite electrodes ("home-made" or commercial ones) to the electrochemistry of DNA allowed the electrochemical reduction of various nucleosides [36, 37] and oligonucleotides [38] in acidic media, as well as electrochemical reduction of low-molecular-weight dsDNA [39, 40] and dGMP in neutral media [40]. Despite BDDEs being considered a biocompatible material, scientific papers dealing with the electrochemistry of NAs at BDDEs are rather scarce and rely exclusively on oxidation with no mention of electrochemical reduction of any type of natural NAs or its components at corresponding surfaces (either H-terminated or O-terminated) [41-46]. The fundamental benefit of reducing the selected analytes for the electrochemistry of DNA lies within the monitoring of the DNA/nucleotide-drug interactions or DNA/nucleotide damaging events as most of the oxidation signals of DNA/nucleotides resp. most organic compounds' oxidation signals would be prone to overlapping in the anodic region, and exact evaluation of these events would be hindered. Nevertheless, DNA/nucleotide cathodic signals present at very negative potentials would most likely not be complicated by this problem, and consequent evaluation of these events could be performed.

The main objective of the presented paper is the examination of the possibility of the electrochemical reduction of several selected purine nucleotides – Guanosine 5'-monophosphate (GMP), 2'-Deoxyguanosine 5'-monophosphate (dGMP), Adenosine 5'-monophosphate (AMP), Adenosine 5'-diphosphate (ADP), Adenosine 5'-triphosphate (ATP), pyrimidine nucleotides – Cytidine 5'-monophosphate (CMP), Thymidine 5'-monophosphate (TMP) nucleotides, and low-molecular-weight double-stranded DNA (dsDNA) derived from salmon sperm on "H-terminated"/"O-terminated" pristine polycrystalline boron-doped diamond electrodes either microcrystalline (B-MCDE) or ultranocrystalline (B-UNCDE) at very negative potentials in neutral media. The effect of ambient oxygen on the electrochemical reduction of nucleotides/dsDNA at the surface of B-UNCDE/B-MCDE is evaluated in terms of repeatability, availability, and magnitude of the voltammetric signals of the selected analytes. Moreover, a thorough examination of the advantages/disadvantages of the proposed sensors for the voltammetric determination of investigated compounds in the presence of HER (demonstrated *via* electrochemical reduction of AMP) is also presented.

## 2. Experimental
### 2.1 Chemical vapor deposition of micro-/ultrananocrystalline boron-doped diamond films (B-MCD/B-UNCD)

The films were deposited on Si (111) substrates with a thickness of 280 μm. Firstly, the substrates were ultrasonically nucleated for 40 minutes in a suspension of nanodiamond powder (5 nm in size) in deionised water. The preparation of boron-doped diamond films involved their growth in a linear antenna microwave chemical vapour deposition (LA MWCVD) reactor. The utilised reactor had a maximum power of 2 × 3 kW in pulse mode. For the growth of B-MCD films, trimethyl borate (TMBT) was used as the source of carbon, boron, and oxygen. The gas mixture was $H_2$/TMBT/$CO_2$ with a $CO_2$ to $H_2$ concentration of 0.2%. The deposition pressure was set at 30 Pa, and the substrate temperature was maintained at 600 °C for all experiments. Methane was added to the gas mixture to decrease the crystal size and grow B-UNCD films. The gas mixture used for B-UNCD films was $H_2$/TMBT/$CH_4$/$CO_2$, with a methane concentration of 1% and $CO_2$ of 0.2% with respect to $H_2$. The flow of evaporated TMBT was set at 2% for UNCD film (~3 μm in thickness) and 1% for MCD film (~2.5 μm in thickness), resulting in a B/C ratio of 278,000 ppm and 312,500 ppm, respectively. All process parameters used for the growth of B-MCD and B-UNCD material are summarized in the **Table 1**.

**Table 1** Process parameters used for the growth of B-MCD and B-UNCD by LA MWCVD.

| Sample | TMBT (sccm) | $H_2$ (sccm) | $CO_2$ (sccm) | $CH_4$ (sccm) | $T_S$ (°C) | $P_{MW}$ (kW) | $p$ (Pa) | $t$ (h) |
|---|---|---|---|---|---|---|---|---|
| **B-MCD** | 5 | 500 | 1 | - | 600 | 6 | 30 | 30 |
| **B-UNCD** | 10 | 500 | 1 | 5 | 600 | 6 | 30 | 30 |

sccm - standard cubic centimetres per minute; $T_S$ - substrate temperature; $P_{MW}$ - provided microwave power; $p$ - gas mixture pressure; $t$ - deposition duration

## 2.2 Fabrication of pristine micro-/ultrananocrystalline boron-doped diamond electrodes (B-MCDEs/B-UNCDEs)

BDD films were deposited on $10 \times 20$ mm$^2$ Si substrates. The electrochemically active and contact areas were fabricated using a masking technology. The active surface was fabricated at the bottom of the BDD substrate with a circle shape of 4 mm diameter (12.6 mm$^2$). The electrode contacting area was masked in the upper part of the BDDE of $3 \times 10$ mm$^2$. Following the masking of the active and contact areas of the BDDE, an intrinsic (non-conductive) diamond layer was applied. This provided effective insulation with excellent mechanical and chemical resistance, thereby ensuring the integrity of the substrate.

## 2.3 Reagents and chemicals

Low-molecular-weight double-stranded DNA (dsDNA) derived from salmon sperm, Guanosine 5'-monophosphate disodium salt hydrate (GMP), 2'-Deoxyguanosine 5'-monophosphate sodium salt hydrate (dGMP), Adenosine 5'-monophosphate disodium salt (AMP), Adenosine 5'-diphosphate sodium salt (ADP), Adenosine 5'-triphosphate disodium salt hydrate (ATP), Cytidine 5'-monophosphate disodium salt (CMP), Thymidine 5'-monophosphate disodium salt hydrate (TMP) were all purchased from Sigma-Aldrich, Germany. Stock solutions of dsDNA (5 mg/mL), GMP, dGMP, AMP, ADP, ATP, CMP, and TMP (uniform concentration of $1 \times 10^{-2}$ M) were prepared daily and directly by dissolving an appropriate amount of the substance in a 0.1 mol/L phosphate buffer of pH 7.4 (PB) containing 1 mM KCl as a supporting electrolyte. More diluted solutions of AMP for the voltammetric experiments in the presence of HER were prepared by exact dilution of the stock solution ($2 \times 10^{-2}$ M) with PB.

## 2.4 Apparatus

Boron-doped diamond (BDD) thin film growth was carried out in an LA MWCVD reactor (modified SCIA cube 300, scia Systems Ltd., Germany). Scanning electron microscopy (SEM) images of B-MCDE/B-UNCDE surfaces were obtained using MAIA-3 Field Emission Gun Scanning Electron Microscope (Tescan, Czech Republic). Raman spectra measurements were performed using a Renishaw InVia Reflex Raman microscope (New Mills, UK), equipped with an air-cooled CCD camera and grating with 2400 lines/mm. Voltammetric experiments were carried out using the µAutolab III/FRA2 potentiostat/galvanostat (Eco Chemie, The Netherlands) driven by NOVA 1.11 software (Metrohm Autolab, Switzerland). All measurements were conducted in a three-electrode system using a laboratory-made pristine B-MCDE/B-UNCDE working electrode mounted in a Teflon (PTFE) holder with a platinum contact on BDDE with an electroactive surface diameter of 4.0 mm, an Ag|AgCl|sat. KCl reference electrode, and a platinum foil counter electrode (Elektrochemické Detektory, Czech Republic) in 20 mL glass voltammetric cells at an ambient temperature.

## 2.5 Preparation of the "O-terminated" pristine B-MCDE/B-UNCDE and "H-terminated" pristine B-MCDE

"O-terminated" surface of B-MCDE or B-UNCDE (O-B-MCDE or O-B-UNCDE) was fabricated by immersion of pristine B-MCDE/B-UNCDE into the solution of the blank electrolyte (PB) and immediate application of the highly positive potential of +3.0 V ($E_{dep}$=+3.0 V) for 1 min ($t_{dep}$=1 min). Afterwards, the sensor was rinsed with deionised water for 5 seconds ($t_{rins}$=5 s), once again placed into the solution of blank electrolyte and a single LSV scan from +1.0 to -3.0 V was performed to examine the state of the O-B-MCDE/O-B-UNCDE surface. "H-terminated" surface of B-MCDE (H-B-MCDE) was fabricated by immersion of pristine B-MCDE into the solution of blank electrolyte (PB) and immediate application of the highly negative potential of -3.0 V ($E_{dep}$=-3.0 V) for 1 min ($t_{dep}$=1 min). Afterwards, the sensor was rinsed with deionised water for 5 seconds ($t_{rins}$=5 s), once again placed in the solution of blank electrolyte and a single LSV scan from -0.5 to -3.0 V was performed to evaluate the state of the H-B-MCDE surface. Utilisation of the quotation marks in the case of "O-terminated" B-UNCDE/B-MCDE was chosen on the basis of utilisation of the highly positive potential of +3.0 V which was in contradiction to the utilisation of subsequent semicathodic LSV scan (from +1.0 to -3.0 V) marked by negative potentials strongly related to the HER. On the other hand, in the case of "H-terminated" B-MCDE it was decided on the basis of exposure of B-MCDE to the air oxygen which was in contradiction with the utilisation of corresponding $E_{dep}$ as well as subsequent cathodic LSV scan moving solely in the negative region of potentials (from -0.5 to -3.0 V).

## 2.6 Purging of the analyte solutions with $N_2$ and corresponding voltammetric analysis

At first, all analyte solutions were purged with $N_2$ for a sufficient time of 20 minutes (the volume of the analyte solution was 15 mL; the utilised electrode system was meanwhile placed in the blank electrolyte solution during this event). After this purging period/treatment, renewal of surface (electrochemical activation) of O-B-MCDE/O-B-UNCDE resp. H-B-MCDE was performed. With the finalisation of surface renewal, purging was stopped, and the working electrode was rinsed with deionised water for 5 seconds ($t_{rins}$=5 s) and placed in the analyte solution for 30 s ($t_{inc}$=30 s). After that, the corresponding LSV scan (based on the termination of the surface – O-/H-termination) was performed.

*2.7 Procedures*

In order to analyse the B-MCDE/B-UNCDE films, scanning electron microscopy (SEM) was employed to capture images from both a top-view (0°) and a cross-section view (90°). The main attention was focused on secondary electrons during imaging. By examining the cross-sectional views of fractured samples, the thickness of the deposited diamond films was determined. For Raman spectra measurements, samples were excited by the laser wavelength of 442 nm (Dual Wavelength HeCd laser Kimmon Koha, model IK5651R-G, Japan) with a power of 40 mW. Samples were exposed to excitation light for 50 seconds, and laser light was focused via Leica objective 100×/NA=0.9 to a spot of 1 μm diameter on the sample in a direction perpendicular to the sample plane. After collection, Raman spectra were baseline-corrected using the Assymetric Least Squares smoothing method (assymetric factor E-6, threshold 0.005, smoothing factor 5, 50 iterations) and fitted by Gaussian functions. The optimal experimental parameters for LSV were as follows: a scan rate of 1000 mV/s and a potential step of 2.4 mV in PB. The examination of electrochemical reduction of all selected analytes and repeatability of the results at the surface of O-B-MCDE/O-B-UNCDE and H-B-MCDE involved the performance of triplet of LSV scans (marked by a utilisation of three different approaches). The first LSV scan involved the immersion of the working electrode into the analyte solution and the immediate LSV scan in the corresponding range of potentials. The second LSV scan followed the incubation of the sensor in the analyte solution for the incubation period of 30 s ($t_{inc}$) and the subsequent LSV scan. The third LSV scan was performed after the purging of the analyte solution with $N_2$ for a defined time period according to the protocol explained in *2.6*. All voltammograms were recorded three times (*n*=3). Endpoints of the corresponding LSV recordings were cut off in order to prevent the inclusion of the artefacts to the voltammetric profiles.

3. Results and discussion

*3.1 Morphological and structural characterisation of pristine micro-/ultrananocrystalline boron-doped diamond electrodes (B-MCDEs/B-UNCDEs)*

To ensure accurate measurements, all experiments were conducted using a three-electrode system. The working electrode was a laboratory-made pristine B-MCDE/B-UNCDE, as shown in **Fig. 1-A**. BDD films were deposited on Si substrates in size 10 × 20 mm². The BDD electrode was carefully mounted in a Teflon (PTFE) holder with a platinum contact on the BDDE. To create the electrochemically active and contact areas, a masking technology was employed. The active surface, with a circular shape of 4.0 mm diameter (12.6 mm²), was fabricated at the bottom of the BDDE substrate. The electrode contacting area, measuring 3 × 10 mm², was masked in the upper part of the BDDE. To enhance the protection and integrity of the substrate, an intrinsic (non-conductive) diamond layer was selectively deposited on non-masked areas of the BDDE. This technological approach/methodology yielded effective insulation with exceptional mechanical and chemical resistance.

To gain insights into the surface morphology of the pristine B-MCDE/B-UNCDE films prior to the voltammetric experiments, SEM measurements were employed. The SEM micrographs of the B-MCDE/B-UNCDE samples are shown in **Fig. 1-B/C**. The B-MCDE film (**Fig. 1-B**) exhibits micrometre-sized faceted granular structures consisting of grains in size of approximately 0.5-1 μm. Additionally, the film thickness was determined to be 3 μm. Contrary to the B-MCDE morphology, the B-UNCD film (**Fig. 1-C**) revealed its ultrananocrystalline nature. The grain size of the B-UNCDE film was estimated to be around 20-70 nm, and the film thickness was 2.5 μm. Acquired SEM characterisations provided valuable information about the surface features and composition of the B-MCDE/B-UNCDE films, allowing to better understand their properties and behaviour in connection with subsequent voltammetric experiments. The fundamental electrical characteristics of BDDE films were quantified through the utilisation of a four-point probe (Ossila Ltd., UK), which enabled the measurement of resistivity and material conductivity. The four-point probe head employs spring-loaded, rounded contacts in lieu of sharp needles, thereby ensuring optimal electrical contact without puncturing the sample. The specific resistivity of the B-MCDE resp. B-UNCDE film was found to be 7.6 mΩ cm, resp. 3.7 mΩ cm.

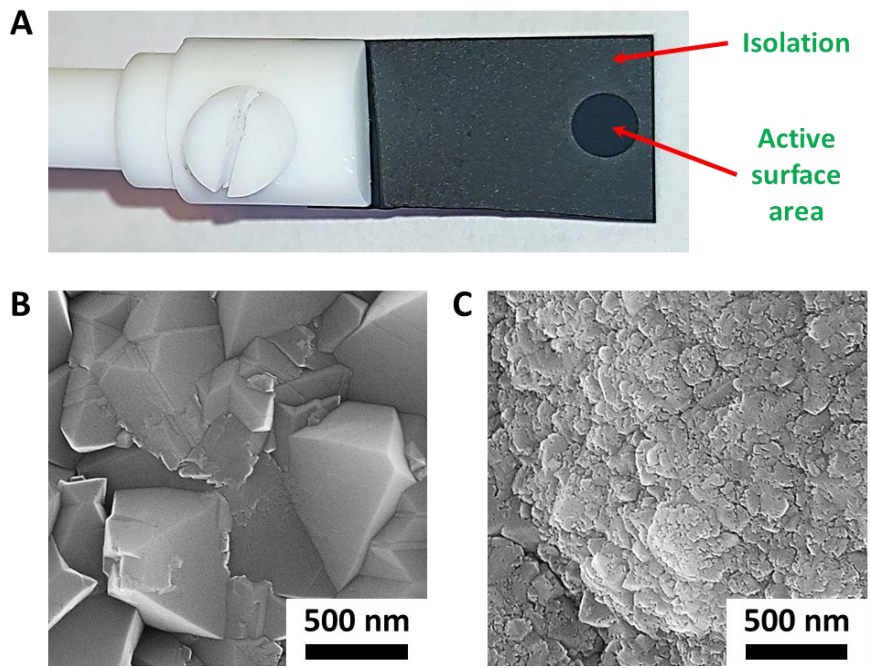

**Fig. 1** Digital photo of BDDEs in PTFE holder with Pt contact (A). SEM images – top view of B-MCDE (B) and B-UNCDE films (C).

The chemical composition of these nanostructures was further investigated to gain a deeper understanding of their properties and behaviour in subsequent voltammetric experiments. The Raman spectra (**Fig. 2-A/B**) reflect well the crystalline character of B-UNCDE and B-MCDE. The broad peaks $B_1$ with the maximum around 460 - 470 cm$^{-1}$ and $B_2$ at the vicinity of 1200 cm$^{-1}$ are all known as characteristics of highly-doped BDD film and find their origin in the breakdown of selection rules due to the size reduction of coherent domains, which allow phonons from high-symmetry points at the zone boundary to contribute to Raman spectra [47]. Here, the reduction of the coherent domain is caused by the incorporation of substitutional boron, which disturbs the translational symmetry of the crystalline structure, by introducing bond angle disorder. Both $B_1$ and $B_2$ are Fano-shaped [48], and after deconvolution of $B_1$ to two separate components ($B_{1a}$ and $B_{1b}$), the positions of $B_{1a}$ can be used to determine boron doping levels of B-UNCDE (at 458 cm$^{-1}$) and B-MCDE (at 478 cm$^{-1}$) to approximately $2.4 \times 10^{21}$ cm$^3$ and $9.2 \times 10^{20}$ cm$^3$, respectively [49]. The peak around 1080 - 1090 cm$^{-1}$ is assigned to trans-polyacetylene and transnano-polyacetylene fragments laying at the grain boundaries and surfaces and $sp^3$ diamond nanocrystallites [50, 51]. The $sp^3$ diamond peak appearing at 1332 cm$^{-1}$ in pure diamond is here red-shifted (ZCP$_D$) to 1300 cm$^1$ and 1280 cm$^1$ for B-MCDE and B-UNCDE, respectively, due to the Fano effect reflecting the boron incorporation into the diamond bulk structure [52-54]. The $sp^2$ carbon band (G-band) appears near 1518 cm$^{-1}$ at relatively low intensity, which highlights the good quality of the BDD films. To emphasize the main differences between B-UNCDE and B-MCDE material, the peak area (integrated intensity) below every single peak and the ZCP$_D$/G band and ZCP$_D$/B$_1$ ratios enabling a qualitative comparison of $sp^3$/$sp^2$ ratio and boron incorporation were calculated and summarized – see **Table S6** and **Table S7**.

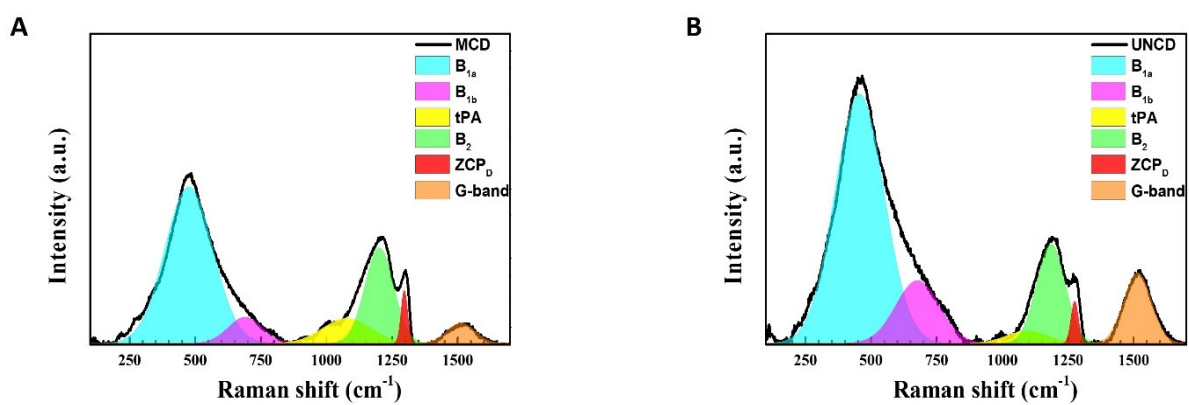

**Fig. 2** Raman spectra for B-MCDE (**A**) and for B-UNCDE (**B**) films.

*3.2 Characterisation of pristine micro-/ultrananocrystalline boron-doped diamond electrodes (B-MCDEs/B-UNCDEs) by means of linear sweep voltammetry (LSV)*

Fabrication of O-B-MCDE/O-B-UNCDE followed the proposed renewal protocol (electrochemical activation) as stated in *2.5* (involving utilisation of the highly positive potential of +3.0 V). In this case, O-B-MCDE is characterised by a larger potential window (**Fig. 3-B**), most likely due to the lower content of $sp^2$ carbon, and corresponding voltammetric profiles are characterised by the appearance of a duplet of voltammetric signals at -0.92 V and -1.36 V (O-B-MCDE) resp. 0.93 V and -1.39 V (O-B-UNCDE) corresponding to the oxygen reduction (**Fig. 3-D**). On the other hand, the fabrication of "H-terminated" B-MCDE involved the utilisation of highly negative potentials (-3.0 V). Moreover, the utilised B-MCDE was exposed to the air aging for 60 days prior to the commencing of the voltammetric experiments as the application of freshly prepared B-MCDE would be characterised by a significantly smaller potential window (**Fig. 3-A**, **red line**) and, therefore, could be utilised, in this case, solely for the electrochemical reduction of CMP (characterised by the lowest $E_p$ for the group of selected analytes) – see **Fig. S1**. Additionally, one of the most pronounced differences between the duplet of electrode surfaces lies within the appearance of a second voltammetric signal at -1.55 V (possibly corresponding also to the oxygen reduction due to the exposure to the air oxygen) as well as the shifting of $E_p$ of the first of the signals from -0.95 V to -1.05 V (also suggesting possible partial oxygenation of "H-terminated" surface; **Fig. 3-C**). The presented contradiction between the relatively broad potential window and high boron content is thoroughly discussed and explained in [55].

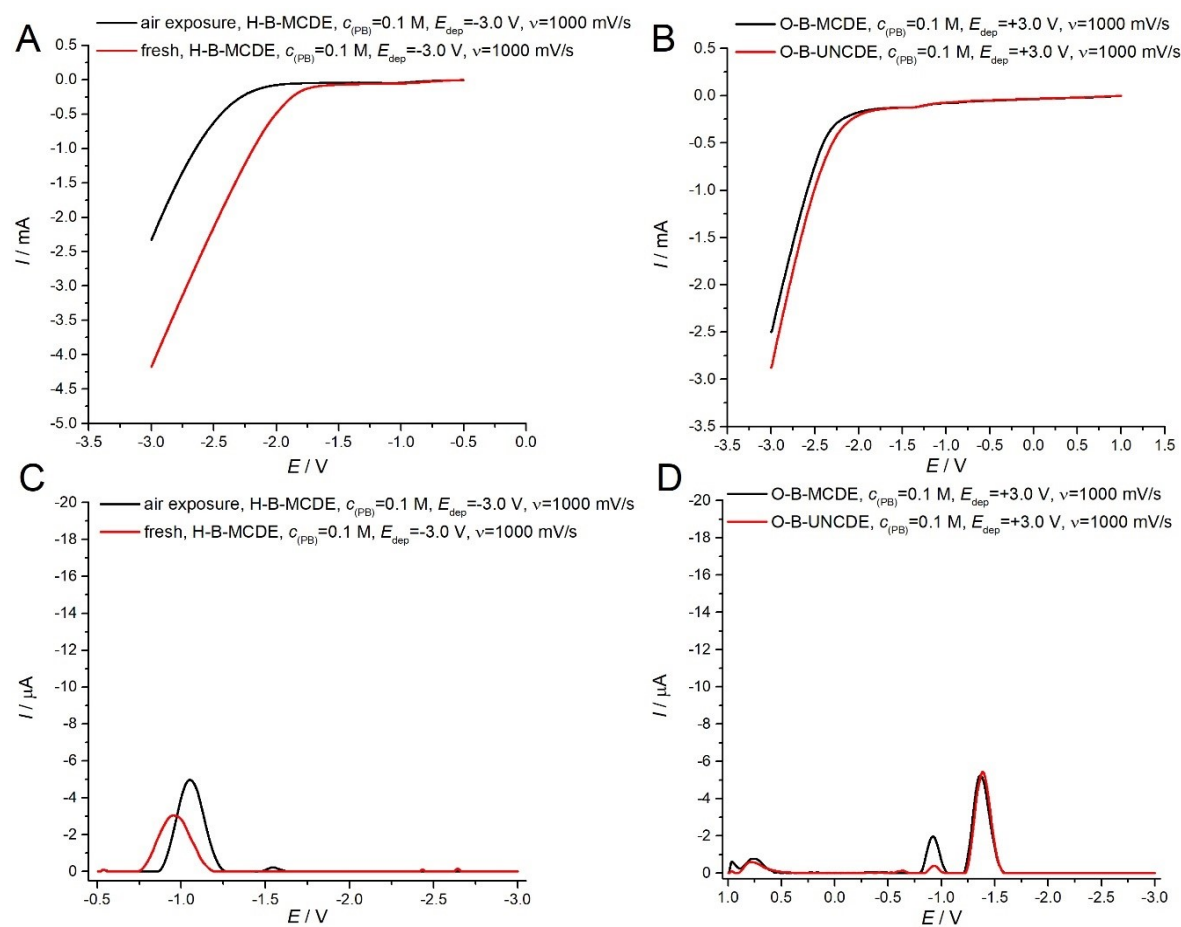

**Fig. 3** LSV (**A**, **B**) and baseline-corrected LSV recordings (**C**, **D**) corresponding to the $O_2$ reduction in the solution of blank electrolyte (PB) at the H-B-MCDE (**A** and **C**) and O-B-MCDE/O-B-UNCDE (**B** and **D**).

*3.3 Electrochemical reduction of purine nucleotides at the O-B-MCDE/O-B-UNCDE resp. H-B-MCDE*

*3.3.1 Electrochemical reduction of the Guanosine 5'-monophosphate (GMP) and 2'-Deoxyguanosine 5'-monophosphate (dGMP) at the O-B-MCDE/O-B-UNCDE resp. H-B-MCDE – probing the effect of the sugar moiety*

The electrochemical reduction of GMP resp. dGMP at the surface of O-B-MCDE/O-B-UNCDE) is marked by the appearance of a single voltammetric signal at -2.36 V/-2.43 V resp. -2.37 V/-2.44 V (**Fig. 4**; $E_p$ for the third LSV scan – **orange line**). In this case, an insignificant difference (in relation to $i_p$ or $E_p$) between the electrochemical reduction of GMP and dGMP was noted (suggesting no pronounced effect of sugar moiety). Main difference between "O-terminated" surfaces was found to be in the $i_p$ of voltammetric signals between the triplet of utilized LSV scans – **red**, **purple** resp. **orange line**, as progressive increase of $i_p$ and shifting of $E_p$ was registered for O-B-UNCDE (**Fig. 4-C/D**), which is contradiction with the trend recorded for O-B-MCDE (**Fig. 4-A/B**) where the utilization of different LSV scans was solely manifested within the enhancement of repeatability of the results – see **Table S2**. This phenomenon could be possibly explained by a higher content of $sp^2$ carbon in the B-UNCD material and therefore higher affinity for the spontaneous adsorption of the selected analyte manifested through the incubation period of 30 s.

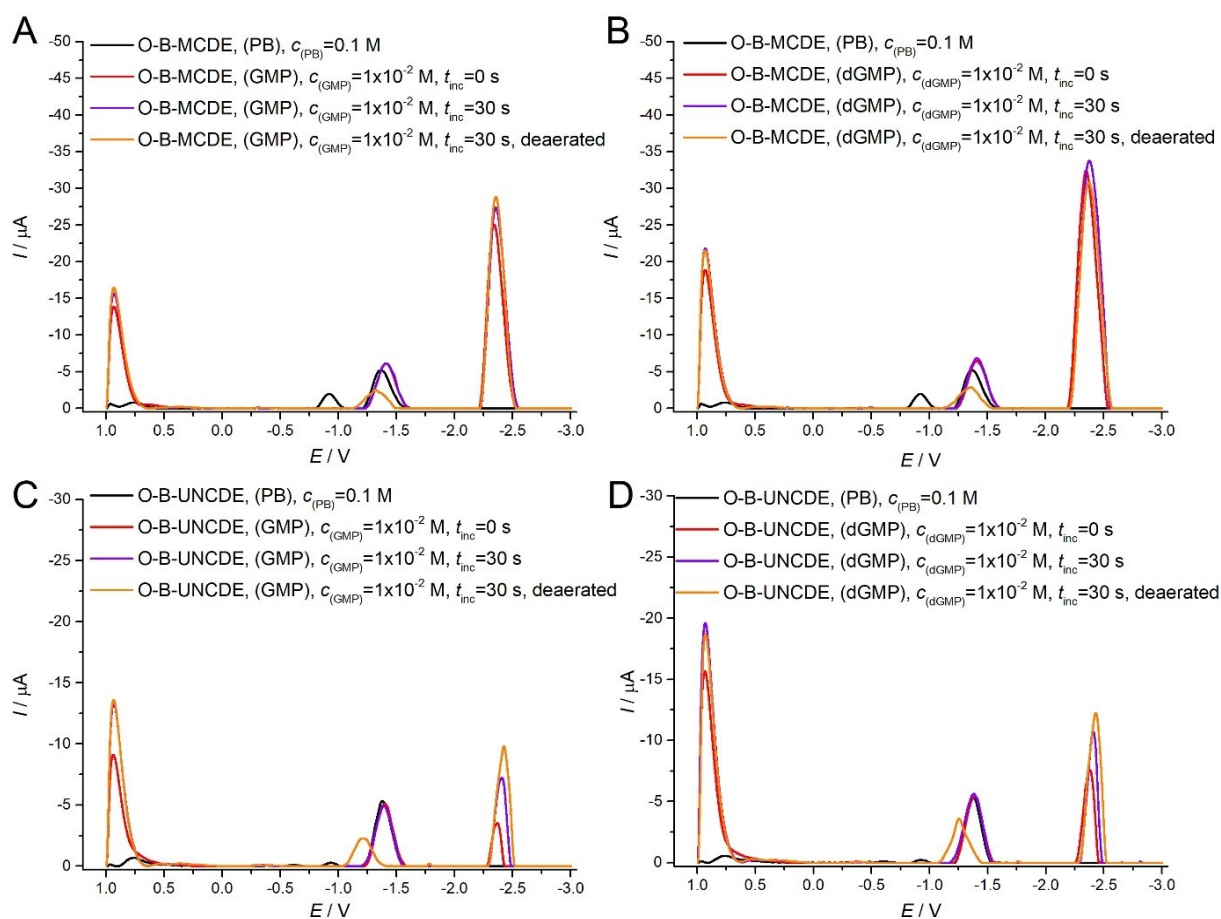

**Fig. 4** Baseline-corrected LSV recordings corresponding to the electrochemical reduction of GMP (**A**), dGMP (**B**) at the O-B-MCDE and GMP (**C**), dGMP (**D**) at the O-B-UNCDE ($c_{(GMP/dGMP)}=1\times10^{-2}$ M).

The electrochemical reduction of the GMP at the H-B-MCDE provided no voltammetric response (**no recordings**). In this case, it can be expected that the electrochemical reduction of the GMP will take place solely on more hydrophilic surfaces. On the other hand, dGMP reduction at the H-B-MCDE was achieved for the second and third LSV scan (marked by a single voltammetric signal at -2.29 V; see **Fig. S2**). Taking into account the third LSV scan (**orange line**), it can be concluded that preceding oxygen reduction plays a pivotal role in dGMP reduction at the surface of the H-B-MCDE (great increase in relation to $i_p$ and repeatability of the results – see **Table S3**). Moreover, when compared to the reduction of the GMP/dGMP at O-B-MCDE/O-B-UNCDE (**Fig. 4, orange line**), it can be observed the absence of the signal corresponding to oxygen reduction for the third LSV scan (see **Fig. S2**, **orange line**) additionally supporting the idea of a facilitation of the GMP/dGMP reduction at more hydrophilic substrates. Based on these observations, it can be also assumed that the presence of sugar moiety (ribose/deoxyribose) plays an important role in the electrochemical reduction of guanine nucleotides at the H-B-MCDE (contrary to the O-B-MCDE/O-B-UNCDE).

*3.3.2 Electrochemical reduction of the Adenosine 5'-monophosphate (AMP), Adenosine 5'-diphosphate (ADP), Adenosine 5'-triphosphate (ATP) at the O-B-MCDE/O-B-UNCDE resp. H-B-MCDE – probing the effect of the phosphate residue*

The electrochemical reduction of AMP at O-B-MCDE/H-B-MCDE is characterised by an appearance of the duplet of voltammetric signals at -2.23 V/-2.22 V or -2.56 V/-2.60 V (**Fig. 5-A/B, red line**). On the contrary, electrochemical reduction of ADP or ATP is characterised solely by a single voltammetric signal at -2.28 V (**Fig. 5-A/B, purple and orange line**). $E_p$ corresponding to the electrochemical reduction of the triplet of adenine nucleotides at O-B-MCDE/H-B-MCDE, as well as an appearance of the second voltammetric signal for the AMP reduction, is in significant contradiction with the electrochemical reduction of adenine (nucleosides or as a part of a CA peak – mixed reduction of cytosine and adenine bases) at pyrolytic graphite electrodes (despite the utilisation of lower scan rates for the LSV scan or utilisation of an acidic [36-38] or even neutral media [39-40]) with an adenine being, in this case, reducible at the potentials close to the reduction of guanine derivatives and for AMP even at more negative potentials (second voltammetric signal of AMP). On the other hand, utilization of the UNCD material was marked by the absence of the second voltammetric signal of AMP as the electrochemical reduction of the AMP, ADP, and ATP at the O-B-UNCDE (**Fig. 5-C**) is, uniformly, marked by an appearance of a single voltammetric signal – at -2.24 V (AMP) or -2.26 V (ADP/ATP).

In the case of O-B-MCDE, no significant effect of the proposed protocol (purging with $N_2$) on the $i_p$ or $E_p$ of voltammetric signals was registered (with the exception of the enhancement of repeatability of the results – see **Fig. S3 and Table S1**). For the B-UNCD material, a significant effect of the proposed protocol was registered solely for the AMP (see

**Fig. S5-A**) with practically no significant effect in relation to the electrochemical reduction of the ADP resp. ATP (see **Fig. S5-B** and **Fig. S5-C**). Similarly to the utilisation of the O-B-MCDE, the proposed protocol secured the enhancement of the repeatability of the depicted results (see **Table S2**). In the case of H-B-MCDE (see **Fig. S4**), the utilisation of the proposed protocol ($t_{inc}$=30 s without $O_2$, i.e. purging with $N_2$ – **orange line**), revealed a more pronounced effect on the voltammetric signals (predominantly in relation to the utilisation of incubation period – **purple line**) in comparison with the alterations of the voltammetric signals registered for the O-B-MCDE (see **Fig. S3**).

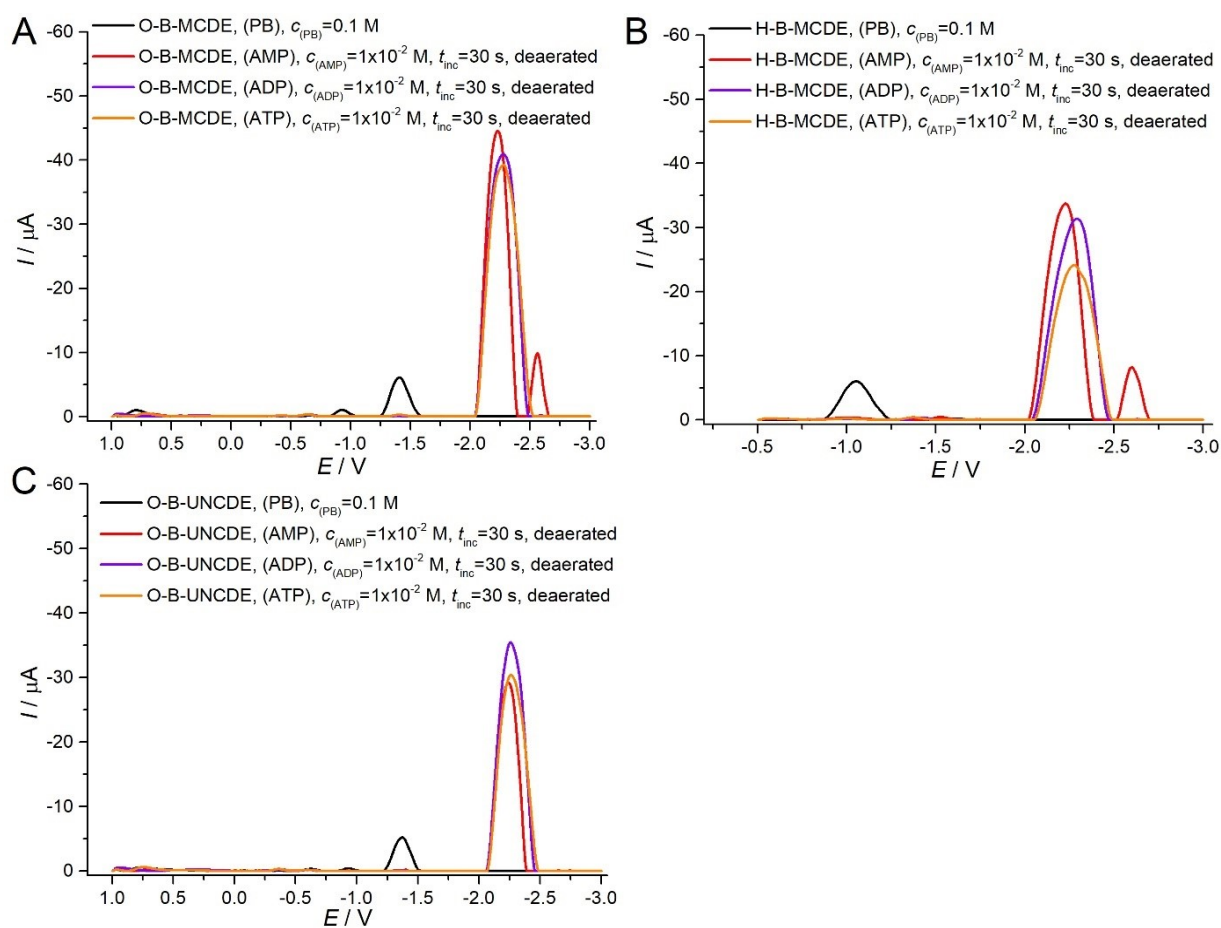

**Fig. 5** Baseline-corrected LSV recordings corresponding to the electrochemical reduction of AMP (**red line**), ADP (**purple line**), and ATP (**orange line**) at the O-B-MCDE (**A**), H-B-MCDE (**B**) and B-UNCDE (**C**) after purging with $N_2$ (third LSV scan; $c_{(AMP/ADP/ATP)}$=1×10$^{-2}$ M).

At last, addressing the comparison for the electrochemical reduction for AMP/ADP/ATP and the third LSV scan, obtained results revealed pronounce shifting of $E_p$ towards negative potentials ($E_{p(AMP)} < E_{p(ADP)} \sim E_{p(ATP)}$) accompanied by a slight progressive decrease of $i_p$ between the individual nucleotides ($i_{p(AMP)} < i_{p(ADP)} \sim i_{p(ATP)}$) at the O-B-MCDE (**Fig. 5-A**) and H-B-MCDE (**Fig. 5-B**). On the other hand, electrochemical reduction of the AMP/ADP/ATP at the O-B-UNCDE (**Fig. 5-C**) was marked by insignificant shifting of $E_p$ (even though the corresponding trend remains preserved – $E_{p(AMP)} < E_{p(ADP)} \sim E_{p(ATP)}$) whereas the trend in relation to $i_p$ is altered as the highest current response was recorded for the electrochemical reduction of ADP ($i_{p(ADP)} < i_{p(AMP)} \sim i_{p(ATP)}$). Observed phenomena can be explained accordingly. In the case of the electrochemical reduction of AMP/ADP/ATP at the surface of B-MCDE (**Fig. 5-A/B**), it is possible to observe a progressive shift of the $E_p$ towards negative potentials with an increase of the phosphate chain length. This can be possibly ascribed to the electrostatic repulsive forces between the negatively charged surface of the transducer (charge imposed by cathodic resp. semicathodic LSV scan during voltammetric analysis) and negative charge of the selected analyte (due to the phosphate residue and its increase with an increase of the phosphate chain length).

On the other hand, this phenomenon was not observed for B-UNCDE (**Fig. 5-C**). This can be presumably ascribed to the the presence of a higher amount of $sp^2$ phase and a lower amount of $sp^3$ phase (present in its structure) suggesting that the adsorption between analyte and transducer being the main driving force, in this case, as opposed to electrostatic forces in the case of the B-MCDE material. Regarding differences in $i_p$, a progressive decrease was observed for B-MCDE material (**Fig. 5-A/B**), this can be possibly ascribed to the above-mentioned electrostatic repulsion between analyte and transducer as the increase of the phosphate chain length could result into bigger repulsion between molecule and transducer surface and, therefore, more difficult accessibility of the nucleobase to the transducer surface for its electrochemical reduction. With B-UNCDE material (**Fig. 5-C**), a different scenario was observed, as the highest $i_p$ was registered for ADP derivative suggesting the adsorption once again being the main driving force of interaction between transducer and analyte (reaching the adsorption maximum for the ADP) over the electrostatic repulsion.

*3.4 Electrochemical reduction of pyrimidine nucleotides at the O-B-MCDE/O-B-UNCDE resp. H-B-MCDE*

The electrochemical reduction of the CMP at the O-B-MCDE is characterised by the occurrence of three voltammetric signals – two overlapping signals (the first signal at -1.88 V and the second signal at -2.06 V) followed by a third voltammetric signal at -2.36 V (**Fig. 6-A**). On the other hand, electrochemical reduction of the CMP at the O-B-UNCDE/H-B-MCDE is marked by an appearance of duplet voltammetric signals, the first one at -1.85 V/-1.77 V and the second one at -2.29 V/-2.37 V (**Fig. 6-B/C**). Nevertheless, due to the low $i_p$ of reduction signals (even after the purging with $N_2$ – **Fig. 6-C**, **orange line**) appearance of a mixed overlapping signal (as in the case of the O-B-MCDE – **Fig. 6-A**) cannot be excluded in the case of O-B-UNCDE. While the presence of single (two overlapping signals) for the electrochemical reduction of cytosine derivatives is in agreement with the former work on the reduction of 2'-Deoxycytidine, resp. various cytosine-containing oligonucleotides at the PGE [37, 38], the appearance of the second (third) voltammetric signal can be solely attributed to the special features of the B-MCDE/B-UNCDE material and its broad potential window (**Fig. 3-B**, **black line**).

Regarding O-B-MCDE (**Fig. 6-A**, **orange line**), the utilisation of the proposed protocol resulted in a significant increase of $i_p$ (in relation to the first two overlapping voltammetric signals), as well as in the significant enhancement of the repeatability of the results at the O-B-MCDE (highest amongst the selected group of nucleotides at the O-B-MCDE – see **Table S1**). On the other hand, the electrochemical reduction of the CMP at the O-B-UNCDE (**Fig. 6-C**, **purple** and **orange line**) was marked by the highest enhancement of repeatability of the results from the selected group of nucleotides, i.e. from 59.3 % (0 s, with $O_2$) to 32.1 % (30 s, with $O_2$) and finally improved to 7.04 % (30s, without $O_2$) (see **Table S2**). At the H-B-MCDE (**Fig. 6-B**, **orange line**), the exceptional effect of oxygen reduction on the reduction of CMP was observed (for both voltammetric signals). At last, the highest $i_p$ for the CMP reduction was noted for the corresponding surface suggesting facilitation of this process at more hydrophobic surfaces. This assumption is also supported by an exceptionally high current response obtained for the electrochemical reduction of the CMP (highest amongst the selected quadruplet of utilised BDDEs surfaces) at freshly prepared H-B-MCDE (see **Fig. S1**). Nevertheless, this can be confirmed solely on the basis of the first voltammetric signal as the second signal is in this case covered by HER.

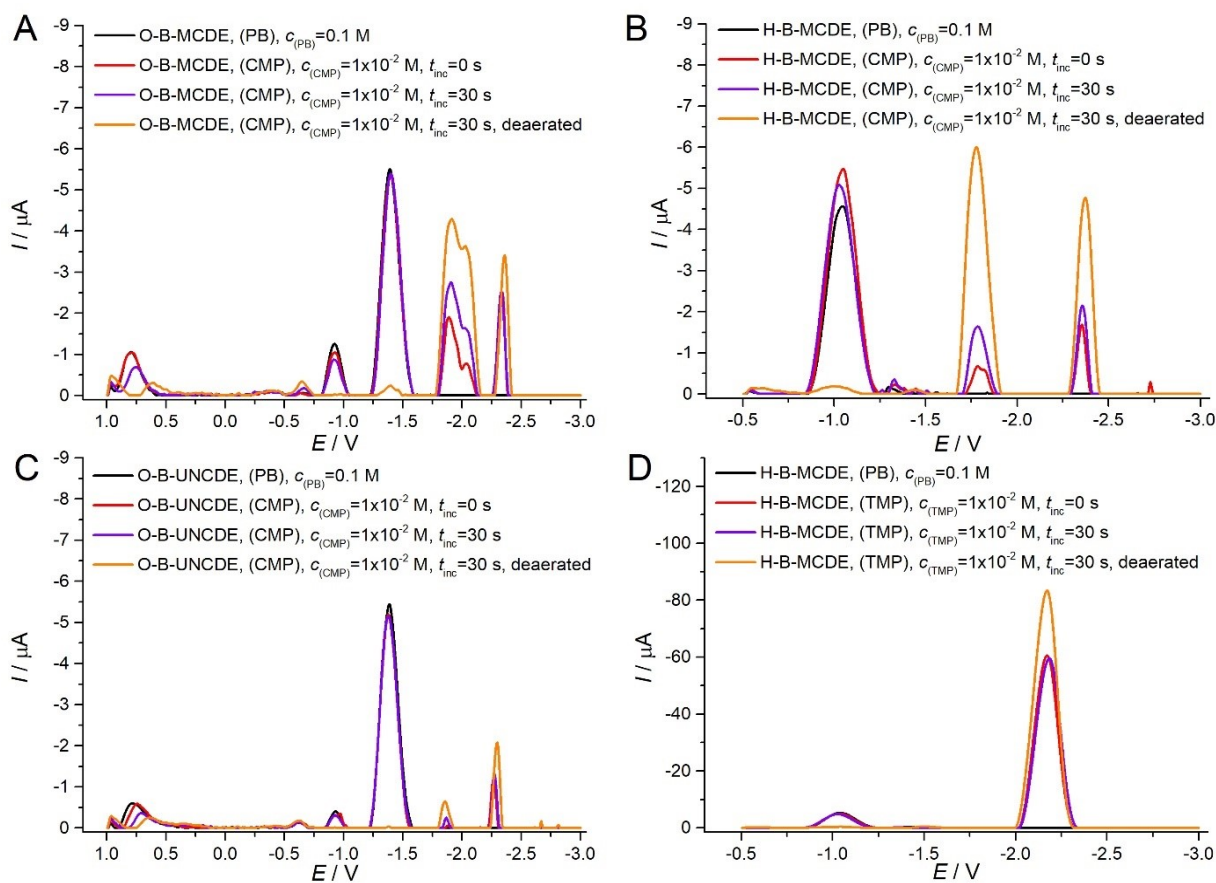

**Fig. 6** Baseline-corrected LSV recordings corresponding to the electrochemical reduction of the CMP at the O-B-MCDE (**A**), H-B-MCDE (**B**), O-B-UNCDE (**C**) and the TMP at the H-B-MCDE (**D**) ($c_{(CMP/TMP)}=1\times10^{-2}$ M).

The electrochemical reduction of the TMP at the H-B-MCDE (**Fig. 6-D**), resp. O-B-MCDE/O-B-UNCDE (see **Fig. S6**) is marked by the appearance of a single voltammetric signal at -2.18 V. This observation is in contradiction with the reduction of 2'-Deoxythymidine at PGE as this process was predominantly obscured, resulted into the shift of HER and, solely after the previous CV scan in corresponding range of potentials, reduction of dTMP was observed [37]. Addressing electrochemical reduction of the TMP at the "O-terminated" surfaces (see **Fig. S6-A/B**), utilisation of the proposed protocol was marked by satisfactory results in terms of repeatability (see **Table S1/2**) with no significant effect on $i_p$ or $E_p$. On the other hand, similarly to the electrochemical reduction of the CMP at the H-B-MCDE – **Fig. 6-B**, exceptional effect of oxygen reduction on electrochemical reduction of the TMP (both in terms oF $i_p$ or $E_p$, as well as repeatability of the results – see **Table S3**) was observed suggesting that the electrochemical reduction of pyrimidine nucleotides at hydrophobic surfaces is significantly affected by the presence of ambient oxygen.

## 3.5 Electrochemical reduction of the double-stranded DNA (dsDNA) at the O-B-MCDE/O-B-UNCDE resp. H-B-MCDE

At last, the possibility of the electrochemical reduction of low-molecular-weight double-stranded DNA (dsDNA) at the surface of O-B-MCDE/O-B-UNCDE resp. H-B-MCDE was examined. Regarding the utilisation of "O-terminated" surfaces, no signal corresponding to the electrochemical dsDNA reduction was observed (**no recordings**) suggesting that dsDNA remains under the utilised conditions electrochemically inactive in relation to the electrochemical reduction. On the contrary, the electrochemical activity of dsDNA at H-B-MCDE (in relation to the electrochemical reduction) was characterised by a single voltammetric signal at -2.04 V (regarding $E_p$ for the LSV scan obtained after the purging with $N_2$; see **Fig. S7**) that can be most likely (despite the $E_p$ for the AMP reduction) attributed to the mixed electrochemical reduction of cytosine and adenine bases which is in correlation with the electrochemical reduction of low-molecular-weight dsDNA at PGE [38-40]. Interestingly enough, the utilisation of the common protocol (involving incubation of the sensor in the analyte solution for 30 s – $t_{inc}$=30 s) resulted, at first, in a significant decrease of $i_{p(dsDNA)}$ ($t_{inc}$=30 s; see **Fig. S7**, **purple line**), as well as in the the negative shift of $E_{p(dsDNA)}$. Moreover, subsequent purging with $N_2$ (see **Fig. S7**, **orange line**) resulted in a significant increase of $i_{p(dsDNA)}$ (the usual phenomenon observed for the redox behaviour of the selected nucleotides). At last, both LSV scans were characterised (apart from the trend observed for the nucleotides) by a progressive decline in terms of repeatability (from 1.88% to 6.95% resp. 7.12% – see **Table S3**). It should be also emphasised that the reactive oxygen species produced by oxygen reduction (prior to the electrochemical reduction of selected analytes) can, to lesser or greater extent, possibly affect the stability of the electrochemical reduction of nucleotides/dsDNA *via* oxidative damage. This extent can be subsequently evaluated through the comparison of $i_p$ obtained for the experiments in aerated and deaerated solutions of the analytes. Nevertheless, the exact effect could be estimated only for pyrimidine nucleotides – TMP and CMP, and dsDNA, since these analytes are electrochemically reduced below the potentials corresponding to the HER region (or at its very beginning).

## 3.6 Voltammetric determination of the Adenosine 5'-monophosphate (AMP) in the presence of HER at O-B-MCDE/O-B-UNCDE resp. H-B-MCDE

In order to examine the effect of the HER on the voltammetric determination of selected analytes, AMP was selected as an ideal candidate being reducible at the most negative potentials (with the exception of GMP and dGMP, respectively) and, therefore, also most strongly exposed to the effect of HER. In this case, examination of the possibility of GMP and dGMP voltammetric determination in the presence of HER was omitted as the electrochemical reduction of the duplet is severely distorted at the H-B-MCDE (see **Fig. S2**). The voltammetric determination of AMP at the O-B-MCDE (**Fig. 7-A**) resp. O-B-UNCDE (**Fig. 7-B**) in the presence of HER was achieved solely in the narrow concentration range of $2\times10^{-2}$ - $2\times10^{-3}$ M and $2\times10^{-2}$ - $4\times10^{-3}$ M, respectively. Determination of the lower concentrations of AMP ($2\times10^{-3}$ M; **Fig. 7-A – blue line**) at the O-B-MCDE can be presumably attributed to the broader potential window – **Fig. 3-B/D**, **black line**. Moreover, voltammetric signals corresponding to electrochemical reduction of AMP below $2\times10^{-3}$ M were not registered, possibly due to the increased effect of HER as the less occupied surface of the sensor (in this case by molecules of AMP) allows the increased hydrogen generation and therefore disturbs the voltammetric determination of the selected analyte. This can be also supported by the shape deformation of the voltammetric signals observed for the lowest concentrations – **olive/blue line**. Nevertheless, the utilisation of the proposed protocol ($t_{inc}$=30 s and purging with $N_2$) for the whole concentration range did not result in the decrease of the repeatability of the results for both utilised sensors in the presence of HER (see **Table S5**).

On the other hand, the voltammetric determination of AMP (*via* its reduction signals) at the H-B-MCDE (**Fig. 7-C**) was achieved solely within the very narrow concentration range of $2\times10^{-2}$ – $6\times10^{-3}$ M. Moreover, the repeatability for the lowest concentration ($6\times10^{-3}$ M) is significantly lower when compared with higher concentrations or with repeatability obtained for the AMP determination at O-B-MCDE/O-B-UNCDE (see **Table S5**). This observed phenomenon can be explained as follows. The utilisation of higher AMP concentrations (($2-1)\times10^{-2}$ M) effectively inhibits the hydrogen evolution reaction (HER) at the H-B-MCDE surface, suggesting saturation by AMP molecules. Although the concentration of water molecules is significantly higher than that of AMP, the size and complexity of AMP molecules (comprising a nucleobase, sugar residue, and negatively charged phosphate moiety) provide numerous adsorption sites on the B-MCDE and B-UNCDE materials. This structural advantage may compensate for the disparity in molecular abundance between AMP and water. Nevertheless, moving to the lower AMP concentrations (<$6\times10^{-3}$ M), a higher contribution of the HER occurs and the fragile equilibrium between the "O-termination" (generated during the air exposure of the electrode prior to the commencing and in between the voltammetric experiments) and "H-termination" (generated during the electrochemical activation of H-MCDE and cathodic LSV scan) is disrupted and moved closer to the "H-termination". This shift towards H-termination is evidenced by the significant narrowing of the potential window (see

**Fig. S8-C, olive line**). All corresponding LSV recordings without baseline-correction can be found in **Supplementary material** – see **Fig. S8-A-C**.

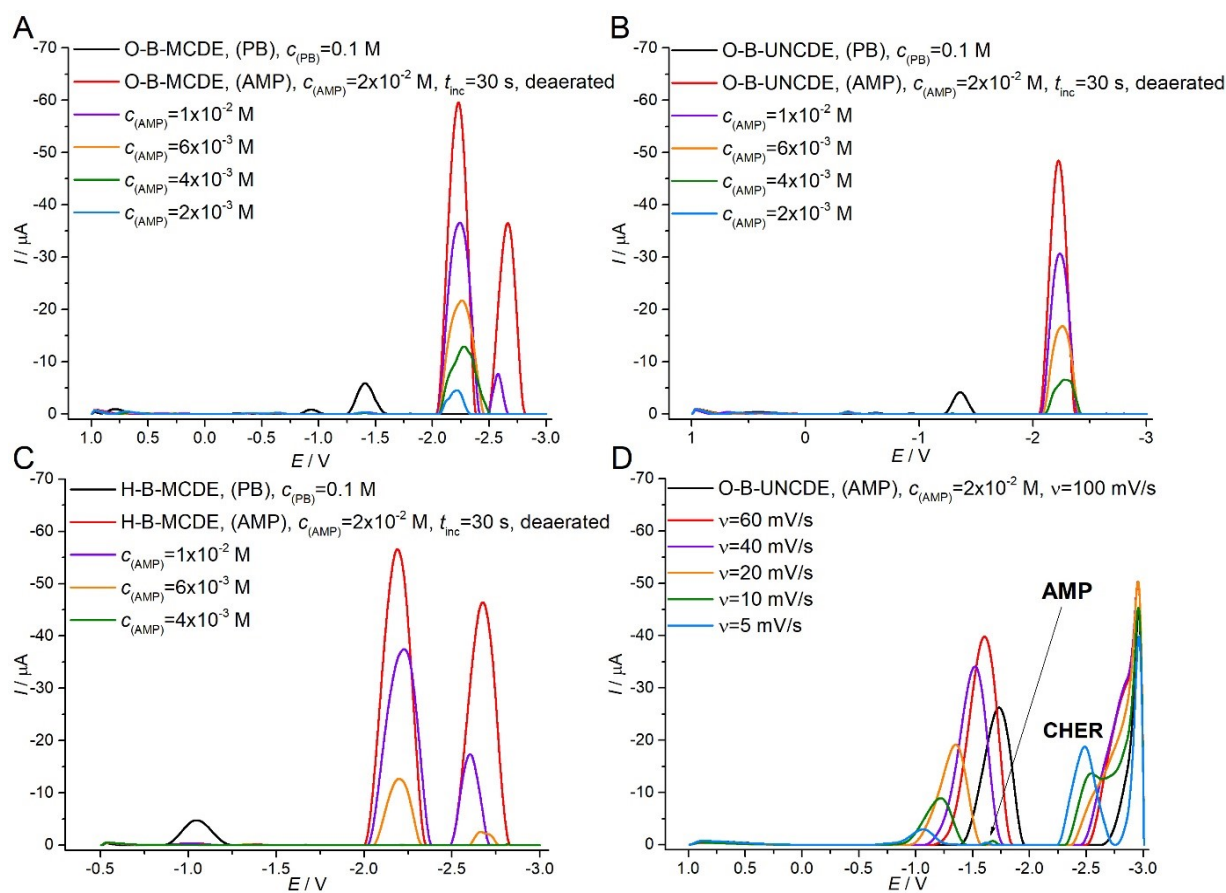

**Fig. 7** Baseline-corrected LSV recordings corresponding to the electrochemical reduction of the AMP in the corresponding range of concentrations at the O-B-MCDE (**A**), O-B-UNCDE (**B**) and H-B-MCDE (**C**) resp. in the range of scan rates of 100 - 5 mV/s with fixed concentration of AMP ($c_{(AMP)}=2\times10^{-2}$ M), $t_{inc}$=30 s and presence of $O_2$ at the O-B-UNCDE (**D**).

One of the main factors affecting to the position of the $E_p$ of the analyte of interest represents the scan rate ($v$). In this case, alternation of this parameter did not result in the desired effect, as for the scan rate from 5000 to 250 mV/s (see **Fig. S8-D**) AMP reduction signal did not move out of the potential region attributed to HER (>-2.0 V). Moreover, subsequent utilisation of even lower scan rates (from 100 to 5 mV/s; **Fig. 7-D**) resulted in an absence of oxygen reduction signal (marked by significant narrowing of the potential window) and also the presence of significant voltammetric signal of the electrode at -1.73 V (**black line**) that can be most likely attributed to higher contribution of HER (due to the extended duration of the LSV scan) and corresponding change of surface morphology. Nevertheless, the AMP reduction signal was observed at very low scan rates (10 mV/s resp. 5 mV/s; **olive** resp. **blue line**), as well as signal corresponding to the catalysed HER (CHER) at -2.48 V possibly due to the extended duration of LSV scan and high concentration of the analyte ($c_{(AMP)}$=2×10$^{-2}$ M). In this case, endpoints of the corresponding LSV recordings were not eliminated in order to demonstrate progressive ascension of CHER. Regarding previous research devoted to the examination of reduction of various oligodeoxynucleotides or ssDNA and its associated catalytic effect at mercury/amalgam electrodes [57, 58], it can be concluded that even the presence of single nucleotide – AMP, in this case, could catalyse HER (to some extent) at B-UNCDE in neutral pH at scan rates below 100 mV (**Fig. 7-D**) suggesting slow catalytic process as previously seen at the HMDE. Despite this finding, further research needs to be done in this manner as the uniform scan rate of 1000 mV/s utilised in this work (used primarily for the determination of the selected group of analytes in the presence of HER with no emphasis put on their catalytic effect) could not fully evaluate on the presence/absence of CHER for the dsDNA and other nucleotides (with an exception of AMP) and its catalytic effect at B-UNCDE.

Based on these observations, it can be concluded that the H-B-MCDE represents an inadequate electrochemical sensor (progressive narrowing of the voltammetric potential window) in terms of voltammetric determination of the selected analytes under the utilised conditions (except for very high concentrations of the analytes) and avoidance of the HER region for the utilised voltammetric scan is required in this case. Nevertheless, utilisation of the O-B-MCDE/O-B-UNCDE revealed satisfactory results in connection with the electrochemical reduction and also voltammetric determination (if the contribution of HER is avoided/suppressed) of the selected group of nucleotides along with the great antifouling properties of these sensors towards the very high concentrations of the selected analytes (even just with the utilisation of simple electrochemical cleaning methodology). All these findings have proven that the utilisation of pristine micro-/ultrananocrystalline BDDEs has a great future perspective for the electrochemistry of DNA (monitoring of oxidation as well as reduction processes), especially for electrodes with high boron content and low $sp^2$ content which both will result in high conductivity of the material and the suppression of the HER and/or oxygen evolution reaction (OER). Last but not least, the great antifouling properties of the utilised BDD electrodes (B-MCDE/B-UNCDE) need to be emphasised as the voltammetric determination of the high concentrations of the selected analytes resulted in the great repeatability of the voltammetric experiments suggesting minimal proneness to passivation of the working electrodes.

## 4. Conclusions

In this study, the successful electrochemical reduction of the selected analytes at the surface of B-MCDE/B-UNCDE was achieved. The utilisation of different types of the termination of B-MCD/B-UNCD electrodes ("O-termination"/"H-termination") revealed a remarkable feature – low-molecular-weight dsDNA is not electrochemically reducible at O-B-MCDE/O-B-UNCDE, while GMP cannot be reduced at H-B-MCDE under the utilised conditions. Moreover, the electrochemical reduction of AMP/CMP nucleotides revealed novel voltammetric signals in the far region of HER, previously unreported at any carbonaceous electrode materials. The utilisation of the proposed protocol (involving incubation of the corresponding sensor in the analyte solution for 30 s along with the purging of the analyte solution with $N_2$) yielded the satisfactory results in terms of repeatability (<5% in the vast majority of the cases). The most profound effect of ambient oxygen on the electrochemical reduction of the selected analytes was observed within the reduction of dGMP at H-B-MCDE and CMP at any of the examined electrode surfaces (O-B-MCDE/O-B-UNCDE and H-B-MCDE). Regarding the utilised electrode material (MCDE or UNCDE) and the type of the surface termination ("O-termination"/"H-termination"), the most profound effect of ambient oxygen on the electrochemical reduction of the selected analytes was found at the H-B-MCDE whereas the lowest effect was registered for the O-B-MCDE (apart from an enhancement of the repeatability of the results).

It is obvious that simultaneous determination of all examined nucleotides represents a challenge due to the strong overlapping of the voltammetric signals of purine nucleotides (GMP, dGMP, AMP, ADP, and ATP) at investigated materials. A similar challenge was observed in the study of the reduction of several oligonucleotides showing strong overlapping for the voltammetric signals of cytosine and adenine bases at mercury/pyrolytic graphite electrodes. A possible solution includes the use of elimination voltammetry [56]. We are aware that high concentrations of investigated analytes were used. It was necessary to partially eliminate the effect of HER and observe the reduction of the nucleotides. Nevertheless, we believe that the utilisation of much lower concentrations at B-MCDE/B-UNCDE is possible if the appropriate HER suppression techniques are applied, or the experimental conditions are further optimised. This will be investigated in our further research following this pilot study.

The voltammetric determination of the selected analytes in the HER region (demonstrated *via* electrochemical reduction of AMP) was predominantly achieved for very high concentrations of the analyte, with the lowest concentration of $2\times10^{-3}$ M registered at O-B-MCDE indicating the significant effect of HER on the monitored event. The H-B-MCDE proved unfavourable for the AMP determination because the utilisation of AMP concentrations below $6\times10^{-3}$ M resulted in the narrowing of the potential window and transfer of the surface morphology of the sensor towards more hydrophobic nature. Even though dsDNA detection is often successful at anodic potentials, the possibility of using cathodic potentials is well-established from the pioneering work of Emil Paleček based on mercury electrodes. This study paves the way for substitution of toxic mercury with more environmentally friendly BDD-based materials. In conclusion, the utilised BDDEs (B-MCDE/B-UNCDE) show promising prospects for the electrochemistry of DNA, particularly in connection with hydrophilic surfaces (O-termination) that exhibit antibiofouling properties, fast electrochemical cleaning/renewal of the fabricated sensors and wide potential window for the electrochemical reduction of dsDNA/nucleotides.

## Data availability

The data that support the findings of this study are openly available at
https://doi.org/10.5281/zenodo.13354870.

## Acknowledgements

This work was supported by the Czech Science Foundation (GAČR) grant number 23-04322L (PL no. 2021/43/I/ST7/03205). We acknowledge CzechNanoLab Research Infrastructure supported by MEYS CR (LM2023051) and the OP JAC financed by ESIF and the MEYS SENDISO - CZ.02.01.01/00/22_008/0004596. All authors thank R. Jackiv for SEM measurements.